\documentclass[12pt, draftclsnofoot, onecolumn]{IEEEtran}

\usepackage{ifpdf}
\usepackage{cite}

\ifCLASSINFOpdf
   \usepackage[pdftex]{graphicx}
\else
\fi
\usepackage{array}
\usepackage{url}
\usepackage{epstopdf}
\usepackage{amsmath}
\usepackage{algorithm}
\usepackage{algpseudocode}
\usepackage{multicol}
\usepackage{amssymb}
\usepackage{amsfonts,graphicx,amsthm, lscape}
\usepackage[font=small]{caption}
\usepackage[font=footnotesize]{subcaption}
\usepackage{color}
\usepackage{mathtools}
\usepackage[subnum]{cases}
\usepackage{empheq}
\usepackage{cleveref}
\usepackage{color,amsmath}
\usepackage{booktabs}
\usepackage{graphicx}

\newcommand{\rev}[1]{{\color{black}#1}}

\ifCLASSINFOpdf
\else
\fi
\begin{document}
%
%
\title{Uplink Resource Allocation for Multiple Access Computational Offloading (Extended Version)}
\author{{Mahsa Salmani and Timothy N. Davidson}
\thanks{This work was supported in part by the Natural Sciences and Engineering Research Council
of Canada under grant RGPIN-2015-06631. The authors are with the Department of Electrical and Computer Engineering, McMaster University, Hamilton, Ontario, Canada. Email: \{salmam, davidson\}@mcmaster.ca. A preliminary version of portions of this paper will appear in \textit{Conf.\ Rec.\ 52\textsuperscript{nd} Asilomar Conf.  Signals, Syst. Comput.}, Oct 2018.}
}

\maketitle

\begin{abstract}
\rev{The mobile edge computing framework offers the opportunity to reduce the energy that devices must expend to complete computational tasks. The extent of that energy reduction depends on the nature of the tasks, 
 and on the choice of the multiple access scheme. 
In this paper, we first address the uplink communication resource allocation for offloading systems that exploit the full capabilities of the multiple access channel (FullMA).}
For indivisible tasks we provide a closed-form optimal solution of the energy minimization problem when a given set of users with different latency constraints are offloading, and a tailored greedy search algorithm for \rev{finding} a good set of \rev{offloading} users. For \rev{divisible} tasks we develop a low-complexity
algorithm to find a stationary solution. \rev{To highlight the impact of the choice of multiple access scheme,} we also \rev{consider the 
TDMA scheme, which, in general, cannot exploit the full capabilities of the channel, and} we develop low-complexity optimal  resource allocation algorithms for 
\rev{indivisible and divisible tasks} under that scheme. \rev{The energy reduction facilitated by FullMA is illustrated in our numerical experiments. Further, those results show that} the proposed algorithms outperform existing algorithms in terms of energy consumption and computational cost.
\end{abstract}


\IEEEpeerreviewmaketitle

\section{Introduction}
\IEEEPARstart{T}{he} rapid development of mobile device technology and wireless communication networks is bringing the vision of ubiquitous computing to fruition, at least for tasks of modest complexity. However, as the demand for ubiquity in computationally-intensive and latency-sensitive tasks increases, the limited computation, memory and energy resources of mobile and other small scale devices present significant impediments to progress. 
The Mobile Edge Computing (MEC) framework seeks to address these impediments by offering the devices the opportunity to offload (a portion of) their computational tasks to a local shared computational resource. This offloading option enables the users to execute more computationally complex applications within a certain deadline, and can also prolong the battery lifetime of the devices \cite{mao2017survey, hu2015mobile, mach2017}. 

In order to 
exploit the opportunities provided by the MEC framework, a computational offloading system must address a number of challenges, including the energy that each user would expend to offload (a portion of) its computational task to the access point \cite{barbera2013offload}, the latency requirements of the tasks \cite{lei2013challenges}, contention for the limited communication resources \cite{mao2017survey}, and, in some cases, contention for the limited shared computation resources at the access point \cite{liu2013gearing}. \rev{In order to address those challenges, the available resources must be effectively allocated to the users. The resource allocation problem, which usually targets the energy consumption of the users while ensuring that the latency constraints of the tasks are met, can be formulated as a joint optimization problem over the available communication and computation resources \cite{sardellitti2015joint, munoz2015optimization, munoz2014energy, salmani2016multiple, wang2017optimized, wang2017joint, chen2018resource, salmani_MACO_arxiv}.
}


\rev{The predominant factor in determining the structure of that optimization problem is the nature of the users' computational tasks.}
\rev{Two classes of tasks that are widely considered in the literature are ``indivisible tasks'' and ``divisible tasks'' \cite{khan2014, mach2017mobile}. A task is indivisible if its components are tightly coupled. Such a task must be either completely offloaded or executed locally, e.g., \cite{kumar2010cloud, wu2013tradeoff, sardellitti2015joint}. On the other hand, a divisible task has independent or loosely coupled components, and can be partitioned. Hence, the mobile device can benefit from the implicit parallelism between the access point and the device by offloading a portion of the task while the remainder is executed locally, e.g., \cite{zhang2013energy, wang2016mobile}. Accordingly, the resource allocation problem is structurally different in the cases of binary offloading (for indivisible tasks) and partial offloading (for divisible tasks). We will address the cases of indivisible and data-partitionable divisible \cite{wang2016mobile} tasks in this paper.} 

In a multi-user offloading system, \rev{irrespective of whether it is binary offloading or partial offloading,} the choice of the multiple access scheme can have a significant impact on the energy consumption, especially when the latency constraints 
are tight. In most of the previous work on such systems, the multiple access schemes employed by the system have been restricted \rev{to schemes that are simple, but are unable to exploit the full capabilities of the channel. Those schemes include Time Division Multiple Access (TDMA) and (Orthogonal) Frequency Division Multiple Access ((O)FDMA), which avoid interference by allocating orthogonal channels  to the users, in time and frequency, respectively, and independent decoding, in which the receiver treats interference as noise.} For example, \rev{in \cite{sardellitti2015joint} and \cite{wang2017joint} the energy minimization problem in a TDMA-based multi-user offloading system is considered for binary and partial offloading cases, respectively. An FDMA-based partial offloading system is considered in \cite{mao2017stochastic}. The energy minimization problems for TDMA and OFDMA-based multi-user partial offloading systems were addressed in \cite{you2017energy}, and the corresponding problem for independent decoding was addressed in \cite{sardellitti2015joint}.}
All of those multiple access schemes limit the range of the rates at which the users can operate reliably, and hence the optimal energy consumption cannot be obtained. 

To address those limitations, the main focus of this work is to find the \rev{optimal} user energy consumption of a $K$-user 
 \rev{(binary or partial) offloading system that employs a multiple access scheme that 
exploits the full capabilities of the multiple access channel. That is, a scheme that enables reliable operation at rates that approach the boundary of the capacity region. Examples of such schemes include Gaussian signalling with joint decoding, and Gaussian signalling with optimally-ordered sequential decoding and time sharing \cite{cover2012elements, elgamal1980multi}. For simplicity, we will refer to any such scheme as a ``full'' multiple access (FullMA) scheme.}
\rev{For any FullMA scheme,} we will provide efficient algorithms for optimally allocating the available communication resources to the $K$ users, each of which wishes to complete its computational task within its own specific deadline, either locally or by offloading (a portion of) the task to an access point that has substantial computation resources. 
We will consider this problem for both indivisible and divisible computational tasks.
For the indivisible case, the combinatorial structure of the binary offloading problem of deciding which users will offload their tasks and which will complete them locally suggests a natural decomposition into an outer search strategy for the offloading decisions and the inner optimization of the communication resources for given offloading decisions. That inner subproblem will be referred to as the ``complete offloading'' problem. For the case of partial offloading of data-partitionable divisible tasks \cite{wang2016mobile}, the fraction of each task to be offloaded will be optimized jointly with the communication resource allocation.

Our strategy for solving the resource allocation problem for systems with a FullMA scheme is based on the insights developed in our previous work on the two-user case \cite{salmani2016multiple, salmani_MACO_arxiv}, which suggests algebraic decompositions of the problem. We will exploit the polymatroid structure of the capacity region of the multiple access channel (see \cite{Tse737513}) \rev{in both the complete offloading and partial offloading cases. In the complete offloading case, we will obtain closed-form optimal solution for the energy minimization problem.}
That solution also forms the core of a tailored greedy search algorithm for good solutions to the binary offloading problem.
In the partial offloading case, \rev{our decomposition strategy enables us to obtain closed-form solutions for some of the design variables and to obtain a stationary solution of the energy minimization problem by employing  
a simple coordinate descent algorithm over the $K$ remaining variables.  }

\rev{To highlight the impact of the choice of the multiple access scheme on the energy consumption of an offloading system, 
we will also address} the energy minimization problem for the TDMA scheme. We will show that that problem can be written as a jointly convex $K$-dimensional optimization problem.
In our simulation results, we will show that although there are scenarios in which TDMA provides good performance, there are others in which exploiting the full capabilities of the multiple access channel enables a substantial energy consumption reduction.

A special case of our total energy minimization problem for a $K$-user offloading system with a full multiple access scheme appears in \cite{ding_NOMA_arxiv}. In \cite{ding_NOMA_arxiv} it was assumed that the latency constraints of all the users are the same, while we consider the more general case in which different users have different latency requirements \rev{and we exploit the maximum allowable latency constraint of each user to reduce the  user energy consumption}. In \cite{ding_NOMA_arxiv}, the solutions for the transmission rates, transmission powers, and the fraction of offloaded bits in the partial offloading scenario are obtained iteratively using a variant of the ellipsoidal algorithm. For a $K$-user system, that algorithm imposes a computational cost of $O(K^3)$ operations per iteration. In contrast, the closed-form optimal solutions for the transmission powers and the fraction of offloaded bits provided by the decomposition-based approach developed herein result in an algorithm whose computational cost is only $O(K \log K)$. Since the class of scenarios for which the proposed algorithm is developed includes the scenario of equal latencies for which the algorithm in \cite{ding_NOMA_arxiv} was developed, our algorithm has the same performance as that in \cite{ding_NOMA_arxiv} in the equal latency case. However, our numerical results, and those in \cite{ding_NOMA_arxiv}, show that the number of iterations required by the corresponding algorithm in \cite{ding_NOMA_arxiv} can be quite large. As a result, in the case of a single-antenna access point the proposed algorithm has a significant computational advantage.

An analogous equal-latency assumption has also been considered for the TDMA scheme in\cite{you2017energy}. An additional difference between the problem in \cite{you2017energy} and that proposed herein is that we have considered the dynamic voltage scaling approach \cite{wang2016mobile} for computation energy management in the mobile devices. This approach guarantees the minimum local energy consumption in the users subject to the latency constraints. In our numerical results, we will show that the energy consumption of the problem formulation proposed for the TDMA case in this paper is significantly lower than the energy consumption of the problem formulation in \cite{you2017energy}. 
\vspace{-5pt}

\section{System Model}
\label{system_model}
We will consider a system consisting of $K$ single-antenna users, each of which has a computational task that is to be executed within its own specific latency constraint, and an access point that is equipped with sufficiently large computational resources that the offloaded tasks can be processed without contention. The offloading users are served over a single time slot by the single-antenna (coherent) receiver at the access point, and the channels between the users and the access point are assumed to be frequency-flat and quasi-static. \rev{We will adopt the conventional discrete-time baseband equivalent model with symbol interval $T_s$. Therefore, if $s_k[n]$ denotes the transmitted signal by the $k^{\text{th}}$ user at the $n^\text{th}$ channel use, and if $h_k$ denotes the channel from the $k^{\text{th}}$ user to the access point, then the signal received at the access point at the $n^\text{th}$ channel use is} 
\begin{equation}
\label{channel_model}
y[n] =\textstyle \sum_{k=1}^K h_k s_k[n] +v[n],
\end{equation}
where $v[n]$ is an additive circular zero mean white Gaussian noise of variance $\sigma^2$.

\rev{In order to explore the impact of the multiple access scheme on the energy consumption of the offloading devices, we will tackle the following generic energy minimization problem
\begin{subequations} 
\label{prob_formula_main} 
\begin{align} 
\min_{\substack{\text{offloading fraction}, \\ \text{communication resources}} } \quad   & \text{Total device energy consumption}\\ 
\label{prob_formula_main_b} 
 \text{s.t.} \quad \quad \quad \quad   & \text{Offloading fraction constraints},\\
 \label{prob_formula_main_c}
 & \text{Latency constraints,\hspace{-20pt}}\\ 
\label{prob_formula_main_d}
& \text{Achievable rate region constraints. \hspace{-20pt}} 
\end{align} 
\end{subequations}
The constraints on the fraction of the computational task that is offloaded by each user are determined by the nature of the tasks that the users seek to offload. If the tasks are indivisible, the offloading fraction for each user is either zero (local execution) or one (complete offloading of the task). Alternatively, if the tasks are data-partitionable divisible tasks, in which a simple-to-describe operation is applied, independently, to different blocks of data \cite{wang2016mobile}, the offloading fraction can be modeled as taking any value in $[0, 1]$. Regarding the constraints on the latencies, we will consider the general case in which each user has its specific latency constraint, independent from the latencies of other users. Finally, the achievable rate region describes the set of rates at which reliable communication can be achieved for a given set of transmission powers; e.g., \cite{cover2012elements}. Different multiple access schemes manage the interference between users in different ways and hence have different achievable rate regions. The capacity region is the convex hull of all achievable rate regions and
we will call any multiple access scheme that can operate reliably at all points in the capacity region a ``full'' multiple access scheme.
}

Now, in order to formulate the \rev{generic energy minimization problem},
let $R_k$ and $P_k$ denote the data rate and power (in units per channel use) employed by user $k$ when it is transmitting, respectively. \rev{In addition, let $\{R_k\}_{k=1}^K$ and $\{P_k\}_{k=1}^K$ denote the sets of transmission rates and transmission powers for all users, respectively. In some cases we will simplify that notation to $\{R_k\}$ and $\{P_k\}$. We will use the generic notation $\mathcal{R}\left(\{P_{k}\}_{k=1}^{K}\right)$ to denote the achievable rate region of a multiple access scheme, and hence the rate region constraint can be written as $\{R_{k}\}_{k=1}^{K} \in \mathcal{R}\left(\{P_{k}\}_{k=1}^{K}\right)$, e.g., \cite{cover2012elements, elgamal1980multi, el2011network}. In specifying that constraint for a particular multiple access scheme, we will assume that the data blocks are long enough for the asymptotic characterization to be valid. Under the asymptotic assumption, the achievable rate region of a FullMA scheme is the capacity region (see \eqref{full_multiple_region} below), and for the TDMA scheme, since each user transmits in a different interval, the rate $R_k$ at which it can reliably communicate during that interval is upper-bounded by the classical single-user capacity expression, e.g., \cite{cover2012elements}.\footnote{Extensions to rate regions for finite block lengths (e.g., \cite{polyanskiy2010channel, jazi2012simpler}) will be guided by the insight developed herein.} }

If $B_k$ denotes the total number of bits describing the task of user $k$, then let $\gamma_k B_k$ define the number of bits offloaded by the $k^{\text{th}}$ user, where $\gamma_k \in \{0, 1\}$ for the binary offloading case, and $\gamma_k \in [0, 1]$ for the partial offloading case. 
Accordingly, the time it takes for user $k$ to offload (the portion of) its task is $ t_{\text{UL}_k} = T_s \tfrac{\gamma_k B_k}{R_k}$. 
The energy that it expends in doing so is $\tfrac{\gamma_k B_k}{R_k} P_k$.

\rev{In order to satisfy the latency constraints in \eqref{prob_formula_main_c}, both the offloaded portion of each user's task and the locally retained portion must be completed within that user's specified latency.
To formulate those constraints, we observe that} the structure of data-partitionable tasks is such that the time that it takes for the access point to process the offloaded portion
can be modeled as a simple multiple of its size 
\cite{zhang2013energy},
\begin{equation}
\label{exe_time_offload}
t_{\text{exe}_k} = \delta_c \gamma_k B_k,
\end{equation}
where $\delta_c$ is the time it takes to process one bit at the access point. 
\rev{For indivisible tasks, $\gamma_k \in \{0, 1\}$, and we can use the expression in \eqref{exe_time_offload} 
if we scale $\delta_c$ so that $\delta_c B_k$ is equal to the time that it would take for the access point to complete the task. }

\rev{The time that it takes for user $k$ to communicate (a portion of) the problem to the access point is the sum of any time it has to wait until it can access the channel, $t_{\text{w}_k}$, and the actual offloading time $t_{\text{UL}_k}$. For FullMA schemes each user has immediate access to the channel and hence $t_{\text{w}_k}=0$, whereas for the TDMA scheme users have to wait until their turn; see Section~\ref{Binary_TDMA}.}
If the time it takes for the access point to send the results back to the $k^{\text{th}}$ user is denoted by $t_{\text{DL}_k}$, then the latency constraint of that offloading user can be written as 
\begin{equation}
\label{latency_constraint}
\rev{t_{\text{w}_k}}+t_{\text{UL}_k}+t_{\text{exe}_k}+t_{\text{DL}_k} \le L_k,
\end{equation} 
in which $L_k$ denotes the maximum allowable latency for user $k$. \rev{The time $t_{\text{DL}_k}$ depends on a number of different factors, including the description length of the results of the (partially) offloaded task, which is often considerably shorter than the description length of the task itself. It also depends on the downlink signalling scheme chosen by the access point, and the energy that the access point expends on the downlink. Since our emphasis is on the minimization of the energy expended by the devices (and not the access point) through the selection of a multiple access scheme for the uplink and the corresponding resource allocation, we will model $t_{\text{DL}_k}$ as a (possibly different) constant for each user.  }

The local execution time takes a similar form to that in \eqref{exe_time_offload} when the users employ a conventional computational architecture. Hence, a local latency constraint for data-partitionable tasks takes the form $t_{\text{loc}_k} = \delta_k (1-\gamma_k) B_k \le L_k$,
where $\delta_k$ is the time it takes for the $k^{\text{th}}$ user to process one bit. A \rev{scaling analogous to that after \eqref{exe_time_offload}} can be used for the binary offloading case. 

\rev{To complete the generic formulation, we will let $E_{\text{loc}_k}(\gamma_k)$ denote the energy that user $k$ expends to complete its local computation within its latency constraint. That energy depends on the number of operations that the local processor must perform to complete (the retained portion of) the user's task, and on the energy required to perform each operation. As discussed after \eqref{prob_main_K_user}, the latter depends on the nature of the computational architecture of the device. For an indivisible task, the number of local operations is either zero (when the task is fully offloaded), or a constant (when the task is locally executed). That constant is determined by the complexity of the task. For data-partitionable divisible task, the number of local operations can be modeled as being proportional to the fraction of the description that user $k$ retains \cite{wang2016mobile}. }

Having developed this notation, the generic problem of minimizing the user energy consumption of a system with $K$ offloading users, \rev{which was described in \eqref{prob_formula_main},} can be formulated as
\vspace{-15pt}
\begin{subequations}
\label{prob_main_K_user}
\begin{align}
\label{prob_formula_main_K_user_a} 
\min_{\substack{\{R_{k}\}, \{P_{k}\},  \{\gamma_{k}\}}}  & \textstyle \sum_{k}  \tfrac{\gamma_{k}B_k}{R_{k}}P_{k} + E_{\text{loc}_k}(\gamma_k)\\ 
\label{prob_formula_main_K_user_b}  
 \text{s.t.} \quad \quad & \rev{ \gamma_{k} \in \{0,1\}\text{ or }  \gamma_{k} \in [0,1]}, \quad  \forall k, \\  
 \label{prob_main_K_user_c}
 &\rev{t_{\text{w}_k}}+T_s  \bigl(\tfrac{\gamma_{k}B_{k}}{R_{k}}\bigr)+\delta_c \gamma_k B_k +t_{\text{DL}_k} \le L_{k}, \quad \forall k, \\
  \label{prob_main_K_user_d}
 &   \delta_k (1-\gamma_k) B_k \leq L_k, \quad  \forall k, \\
 \label{prob_main_K_user_e}
& 0 \le P_{k}, \quad  \forall k, \\ 
\label{prob_main_K_user_f}
&\rev{ \{R_{k}\}_{k=1}^{K} \in \mathcal{R}\left(\{P_{k}\}_{k=1}^{K}\right),}
\end{align}
\end{subequations}
\rev{where the constraints in \eqref{prob_formula_main_K_user_b} are the offloading fraction constraints for binary or partial offloading, respectively, \eqref{prob_main_K_user_c} and \eqref{prob_main_K_user_d} capture the latency constraints on the offloaded and locally-executed portions of the task, and \eqref{prob_main_K_user_f} is the rate region constraint for the chosen multiple access scheme.}

\rev{Our primary algorithm development for the solution of \eqref{prob_main_K_user} will be tailored to devices with the dynamic voltage scaling computational architecture \cite{wang2016mobile}. That architecture enables the device to adjust its CPU frequency and hence to minimize the energy it requires to complete (the local portion of) its task within the specified latency constraint. Since in that architecture 
the local latency constraint in \eqref{prob_main_K_user_d} is implicitly satisfied, it can be removed from \eqref{prob_main_K_user}. For a data-partitionable task, the minimized local computational energy can be expressed in the form \cite{wang2016mobile}}
\begin{equation}
\label{loc_opt_energy}
E_{\text{loc}_k} (\gamma_k) = \tfrac{M_k}{L_k^2} \bigl((1- \gamma_{k}) B_k\bigr)^3,
\end{equation}
where the coefficient $M_k$ depends on the characteristics of the chip of user $k$. \rev{For the case of binary offloading with dynamic voltage scaling architecture, we will denote the minimized local energy computation by $\underline{E}_{\text{loc}_k}$, i.e.,  
\begin{equation}
\label{binary_DVS}
E_{\text{loc}_k}(0) = \underline{E}_{\text{loc}_k} \text{and } E_{\text{loc}_k}(1) = 0.
\end{equation}
}

In Sections~\ref{binary_off} and \ref{partial_off} we will focus on the development of algorithms for users that employ dynamic voltage scaling in the binary and partial offloading scenarios, respectively. 
However, with simple modifications the proposed algorithms can be applied to users with conventional computation architectures. The required modifications in the binary case are discussed at the end of Section~\ref{binary_off}, and the modifications for the case of partial offloading were illustrated for a two-user system in \cite{salmaniapcc2017}. In our numerical results in Section~\ref{num_results}, we will illustrate that dynamic voltage scaling approach provides significant energy savings. 

As mentioned in the Introduction, the problem in \eqref{prob_main_K_user} is different from those in \cite{you2017energy} and \cite{ding_NOMA_arxiv}. We allow the latency constraints of the users, $L_k$, to be different, which enables the users with larger latencies to benefit from their own available time to transmit. In \cite{you2017energy} and \cite{ding_NOMA_arxiv} the latency constraints of the users are assumed to be the same, which forces the system to work with the minimum latency constraint among the users. If the latencies are different, doing that will increase the total energy consumption. In addition, for the partial offloading case, the formulations in \cite{you2017energy} and \cite{ding_NOMA_arxiv} assume that $\delta_c$ is small enough that the dependence of the execution time at the access point, $t_{\text{exe}_k}$, on the fraction of the task that is offloaded, $\gamma_k$, can be neglected. We do not make that assumption in our formulations; see \eqref{exe_time_offload}. Finally, in contrast to \cite{you2017energy}, in our formulation we assume that the users can employ dynamic voltage scaling \cite{zhang2013energy, wang2016mobile} to minimize the energy that they expend in local computation. 

\rev{The rest of this paper addresses the energy minimization problem in \eqref{prob_main_K_user} for two classes of computational tasks, namely indivisible tasks and data-partitionable divisible tasks, under two different multiple access schemes, namely FullMA and TDMA. In particular, we will consider that problem in the binary offloading case (for indivisible tasks) under FullMA in Section~\ref{binary_fullMA} and under the TDMA scheme in Section~\ref{Binary_TDMA}. We will tackle the energy minimization problem for a partial offloading system (for data-partitionable divisible tasks) under FullMA in Section~\ref{partial_FullMA}, and under the TDMA scheme in Section~\ref{partial_TDMA}.   }

\section{Binary Offloading}
\label{binary_off}
In this section we will consider minimizing the total energy consumption of the $K$-user system when the computational tasks of the users are indivisible, i.e., the task of each user must be either totally offloaded to the access point or executed by the user. \rev{Since the offloading decision is binary,} the problem of finding the optimal selection of offloading users that
minimizes the total energy consumption
is combinatorial.
As a result, the joint offloading-decision and resource-allocation problem is typically partitioned, with the optimal resource allocation being found for given offloading decisions and a combinatorial search strategy being used to make the offloading decisions. Accordingly, in this section we first seek the optimal solution of energy minimization problem for the complete offloading case in which a subset of users is scheduled to offload their tasks; \rev{see Section~\ref{complete_offloading}.}
Then, \rev{in Section~\ref{greedy_search},} we will develop a low-complexity pruned greedy search technique that is tailored to the characteristics of the problem to find a set of offloading users that typically results in close-to-optimal energy consumption. 

\vspace{-10pt}
\subsection {Complete Computation Offloading} 
\label{complete_offloading}
Let $\mathcal{S} =\{1, 2, \dots, K\}$ denote the set of all $K$ users in the system and let $\mathcal{S}' \subseteq \mathcal{S}$, \rev{where $|\mathcal{S}' | = K'$,} denote the subset of users scheduled to fully offload their tasks, i.e., $\gamma_k = 1, \forall k \in \mathcal{S}'$ and $\gamma_k = 0, \forall k \notin \mathcal{S}'$. \rev{(As mentioned above, the selection of $\mathcal{S}'$ is discussed in Section~\ref{greedy_search}.) In that case, the total device energy consumption consists of the sum of the transmission energies of the users in $\mathcal{S}'$ and the sum of the local computational energies of the remaining users. The latter term can be minimized (while satisfying the latency constraint) by employing optimized dynamic voltage scaling \cite{wang2016mobile}, which leads to the following expression for the total device energy:
\begin{equation}
\label{energy_separate}
E_{\text{total}} = \textstyle \sum_{k \in \mathcal{S}'}  \tfrac{B_k}{R_{k}}P_{k} + \sum_{j \in \mathcal{S} \setminus \mathcal{S}'} \underline{E}_{\text{loc}_j}.
\end{equation}
Thus, the problem that remains is to minimize the energy consumed by the offloading devices 
\begin{subequations}
\label{prob_main_K_user_full_off}
\begin{align}
\min_{\substack{\{R_{k}\}, \{P_{k}\}}}  & \textstyle \sum_{k \in \mathcal{S}'}  \tfrac{B_k}{R_{k}}P_{k} \\ 
\label{prob_formula_main_K_user_full_off_b}  
 \text{s.t.} \quad & \rev{t_{\text{w}_k}}+ T_s  \bigl(\tfrac{\gamma_{k}B_{k}}{R_{k}}\bigr)+\delta_c \gamma_k B_k +t_{\text{DL}_k} \le L_{k}, \forall k \in \mathcal{S}', \\
  \label{prob_main_K_user_full_off_c}
& 0 \le P_{k}, \quad  \forall k \in \mathcal{S}', \\ 
\label{prob_main_K_user_full_off_d}
&\rev{ \{R_{k}\}_{k=1}^{K'}  \in \mathcal{R}\bigl(\{P_{k}\}_{k=1}^{K'}  \bigr).} 
\end{align}
\end{subequations}
}
\rev{As discussed in Section~\ref{system_model}, the achievable rate region, $\mathcal{R}$, and the waiting time, $t_{\text{w}_k}$, depend on the chosen multiple access scheme.} In the following sections, we will provide solutions to \eqref{prob_main_K_user_full_off} for a FullMA scheme and for the TDMA scheme.

\vspace{-10pt}
\subsection {Full Multiple Access Scheme} 
\label{binary_fullMA}
\rev{For a FullMA scheme,
the achievable rate region is the capacity region of the multiple access channel. Since there are $K'$ users in $\mathcal{S}'$, that region can be described by the $K'$ constraints of the form $0 \le R_k$ and the $(2^{K'}-1)$ constraints of the form \cite{cover2012elements, elgamal1980multi}}
\begin{equation}
\label{full_multiple_region}
\textstyle \sum_{i \in \mathcal{N}} R_{i} \le  \log \bigl(1+ \sum_{i \in \mathcal{N}} \alpha_i P_i \bigr),
\end{equation}
in which $\alpha_i = \tfrac{|h_i|^2}{\sigma^2}$ and $ \mathcal{N} \subseteq \mathcal{S}'$. \rev{Furthermore, in a FullMA scheme the available channel is simultaneously assigned to all users, and hence $t_{\text{w}_k} =0, \forall k$. }
Therefore, for a FullMA scheme, the problem in \eqref{prob_main_K_user_full_off} becomes
\begin{subequations}
\label{binary_main_K_user_MAC} 
\begin{align}
\label{binary_main_K_user_MAC_a}
\min_{\{R_{k}\}, \{P_{k}\}}  & \textstyle \sum_{k \in \mathcal{S}'}  \tfrac{B_k}{R_k}P_{k}\\ 
\label{binary_main_K_user_MAC_b}
 \text{s.t.} \quad & T_s  \bigl(\tfrac{B_{k}}{R_k}\bigr) \le \tilde{L}_{k}, \quad \forall k \in \mathcal{S}'\\
 \label{binary_main_K_user_MAC_c}
& 0 \le R_{k},  \quad \forall k \in \mathcal{S}'\\ 
\label{binary_main_K_user_MAC_d}
& 2^{\sum_{i \in \mathcal{N}} R_{i}} \le 1+ \textstyle \sum_{i \in \mathcal{N}} \alpha_i P_i, \quad \forall \mathcal{N} \subseteq \mathcal{S}',
\end{align}
\end{subequations}
where $\tilde{L}_k = L_k - \delta_c \gamma_k B_k -t_{\text{DL}_k}$.

As the first step toward solving the problem in \eqref{binary_main_K_user_MAC}, we decompose the problem into an inner optimization over the transmission powers and an outer optimization over the rates: 
\begin{eqnarray}
\label{binary_main_K_user_MAC_decompose}
& \hspace{-60pt}  \displaystyle{\min_{ \{R_{k}\}}}   & \hspace{10pt} \displaystyle{\min_{\{P_{k}\}}} \hspace{5pt}  \textstyle \sum_{k \in \mathcal{S}'}  \tfrac{B_k}{R_k}P_{k} \\ 
&\quad \quad \text{s.t.}  \hspace{10pt}   \eqref{binary_main_K_user_MAC_b}-\eqref{binary_main_K_user_MAC_d} ,  &\hspace{15pt} \text{s.t.} \hspace{10pt}   \eqref{binary_main_K_user_MAC_d}. \nonumber
\end{eqnarray}
For a fixed set of rates $\{ R_k\}$, the inner optimization problem is a linear programme in $\{ P_k \}$ and the feasibility region for the transmission powers is a polyhedron. Hence, in the search for an optimal solution it is sufficient to restrict attention to the vertices of the feasibility region. Each vertex is described by the simultaneous satisfaction of $K'$ of the linear inequality constraints in \eqref{binary_main_K_user_MAC_d} with equality. As we show in the next section, by exploiting the polymatroid structure of the constraints in \eqref{binary_main_K_user_MAC_d} (e.g., \cite{Tse737513}), we can significantly reduce the number of the candidate vertices. In fact, we will show that we can find a closed-form optimal solution for the powers. 

\subsubsection{Closed-form optimal solutions for the powers}
\label{binary_closed_form_power}
To begin, let us group the rate region constraints in \eqref{binary_main_K_user_MAC_d} into $K'$ classes, where a constraint is assigned to class-$\ell$ if it involves the powers and rates of $\ell$ users; i.e., the constraint is assigned to class $\ell$ if $|\mathcal{N}| = \ell$. In Appendix~\ref{exploit_structure} we show that the vertices of the rate region that are candidates for optimality arise from the simultaneous satisfaction of at most one constraint from each of the classes. Since such vertices involve the simultaneous satisfaction of $K'$ constraints, that implies that at optimality one constraint from each class holds with equality; see also \cite{Mahsaasilomar2018}. 

Since class-$K'$ contains only one constraint, that implies that at optimality 
\begin{equation}
\label{K_th_constraints}
2^{\sum_{i =1}^{K'} R_{i}} = 1+ \textstyle \sum_{i =1}^{K'} \alpha_i P_i.
\end{equation}
Accordingly, the power of any arbitrary user, say user $n$, can be written in terms of the powers of the other users as 
\begin{equation}
\label{n_user_closed_form}
 \alpha_n P_n = 2^{\sum_{i }R_{i}}  - \textstyle \sum_{ i \neq n} \alpha_i P_i -1.
\end{equation}
By substituting this expression
into \eqref{binary_main_K_user_MAC_a} and \eqref{binary_main_K_user_MAC_d}, the inner optimization problem in \eqref{binary_main_K_user_MAC_decompose} remains a linear programming problem, but now with $(K'-1)$ variables, namely,
\begin{subequations}
\label{main_K_user_MAC_powers} 
\begin{align}
\label{main_K_user_MAC_powers_a}
\min_{\{P_{k}\}}  \quad & \textstyle  \sum_{k \in \mathcal{S}' \setminus \{n\}}  \bigr(\rho_k- \rho_n \bigl) \alpha_k P_{k} \\
\label{main_K_user_MAC_powers_b}
 \text{s.t.} \quad & 2^{ \sum_{i \in \mathcal{N}} R_{i}} -1 \le \textstyle \sum_{i \in \mathcal{N}} \alpha_i P_i  \le 2^{R_n} \bigl(2^{ \sum_{i \in \mathcal{N}} R_{i}} -1 \bigr),
\quad \forall \mathcal{N} \subseteq \mathcal{S}' \setminus \{n\},
\end{align}
\end{subequations}
in which 
\begin{equation}
\label{rho_def_app}
\rho_k = \tfrac{B_k}{\alpha_k R_k }.
\end{equation}

It can be seen that the constraints of the problem in \eqref{main_K_user_MAC_powers} have a polymatroid structure, and hence the optimal solution
results from simultaneous satisfaction of $(K'-1)$ constraints with at most one constraint from each class. For positive coefficients of the powers in the objective function in \eqref{main_K_user_MAC_powers_a} it can be shown that, analogous to \eqref{K_th_constraints}, at optimality the single lower bound constraint in class-$(K'-1)$ is satisfied with equality. Accordingly, we can obtain a closed-form solution for the power of another arbitrary user by using an expression analogous to \eqref{n_user_closed_form}.

Based on the above discussion, we can obtain a sequence of closed-form solutions for all the powers for a given set of transmission rates if we can guarantee that in each step all the coefficients $\rho_k- \rho_n$ are positive. We can do that if we determine the permutation $\pi$ so that 
\begin{equation}
\label{ordering_rho}
\rho_{\pi(K')} \le \rho_{\pi(K'-1)} \le \dots \le \rho_{\pi(1)},
\end{equation}
and choose the sequence of values of $n$ to be $\pi(K'), \pi(K'-1), \cdots, \pi(1)$. 
Once the ordering in \eqref{ordering_rho} has been determined, the first step of the algorithm is to obtain the closed-form solution for $ P_{\pi(K)}$ by substituting the expression in \eqref{n_user_closed_form} with $n = \pi(K')$ into \eqref{K_th_constraints}; that is, 
\begin{equation}
\label{K_user_closed_form_iter}
 \alpha_{\pi(K')} P_{\pi(K')} = 2^{\sum_{i=1}^{K'} R_{\pi(i)}} - \textstyle \sum_{ i=1}^{K'-1} \alpha_{\pi(i)}P_{\pi(i)} -1.
\end{equation}
The same procedure can then be applied in a sequential manner to find closed-form solutions for all the powers. In the last step, 
we obtain
\begin{equation}
\label{K_user_closed_form_last_iter}
P_{\pi(1)}= (2^{R_{\pi(1)}}-1)/ \alpha_{\pi(1)}.
\end{equation} 
This expression is only a function of this user's rate and channel, and does not depend on the powers of the other users. By retracing our steps, we obtain a closed-form solution for the optimal power of each user in terms of the rates of other users rather than their powers; i.e.,
\begin{equation}
\label{K_user_closed_form_interms_of_rates}
 P_{\pi(k)} =  \bigl(\tfrac{2^{R_{\pi(k)}}-1}{ \alpha_{\pi(k)}}\bigr) 2^{\sum_{j=1}^{k-1} R_{\pi(j)} }.
\end{equation} 

We observe that the ordering in \eqref{ordering_rho} not only ensures that the terms $(\rho_k- \rho_n)$ in \eqref{main_K_user_MAC_powers_a}, and the corresponding terms in the subsequent instances of \eqref{ordering_rho}, are positive, it also determines the (optimal) decoding order that enables the rates that will be chosen in \eqref{rates_closed_form} below to be achieved by successive decoding. (Since these rates correspond to vertices of the capacity region, no time sharing is required.) In particular, it can be seen from \eqref{K_user_closed_form_last_iter} that 
the message from user $\pi(1)$ is being decoded after the messages from all other offloading users have been decoded and the corresponding interference canceled. Similarly, the expression in \eqref{K_user_closed_form_iter} reveals that the message from user $\pi(K')$ is the first message to be decoded, with the interference from the messages from the other users being treated as noise.

\subsubsection{Closed-form optimal solutions for the rates}
Now that we have the closed-form solutions for the transmission powers in \eqref{K_user_closed_form_interms_of_rates}, the outer optimization problem in \eqref{binary_main_K_user_MAC_decompose} becomes
\begin{subequations}
\label{main_K_user_MAC_rates}
\begin{align} 
\label{main_K_user_MAC_rates_a} 
\min_{\{R_{k}\}} \quad & \textstyle \sum_{k \in \mathcal{S}'}  \tfrac{B_{\pi(k)} }{R_{\pi(k)} }  (\tfrac{2^{R_{\pi(k)} }-1}{ \alpha_{\pi(k)} })2^{\sum_{j=1}^{k-1} R_{\pi(j)} } \\ 
\label{main_K_user_MAC_rates_b}
 \text{s.t.} \quad &   \bigl(\tfrac{T_s B_{k}}{\tilde{L}_k}\bigr) \le  R_k, \quad \forall k \in \mathcal{S}'.
\end{align}
\end{subequations}
It can be shown that the objective function in \eqref{main_K_user_MAC_rates} is an increasing function with respect to each transmission rate and that the constraints on the transmission rates are separable. Hence, the optimal rate for each user is the minimum feasible rate according to its latency constraint, 
\begin{equation}
\label{rates_closed_form}
R_{k} = \tfrac{T_s B_{k}}{\tilde{L}_k}.
\end{equation}

Since this expression depends only on the parameters of the problem, we can obtain the $\rho_k$'s in \eqref{rho_def_app}. Once those $\rho_k$'s have been sorted, the optimal solutions for the transmission powers can be found using \eqref{K_user_closed_form_interms_of_rates}. These steps are summarized in Algorithm~\ref{full_algorithm}. The computational efficiency of the algorithm is apparent from the fact that the number of operations required is dominated by the sorting procedure in Step 3, which requires $O(K' \log K')$ operations. 
\begin{algorithm}[htp]
\begin{small}
\caption{: The optimal solution to \eqref{binary_main_K_user_MAC}} 
\label{full_algorithm}
\begin{algorithmic}
\State \textbf{Input data:} $\mathcal{S}'$, $\{ B_k \} $, $\{ \tilde{L}_k \}$, $\{ \alpha_k \}$, and $T_s$. 
\State \textbf{Step 1:} Calculate the optimal rates $\{ R_k \}$ using \eqref{rates_closed_form}.
\State \textbf{Step 2:} Calculate the values $\{\rho_k\}$ using \eqref{rho_def_app}.
\State \textbf{Step 3:} Order  $\{ \rho_k\}$ according to \eqref{ordering_rho} to find the optimal permutation $\pi$.
\State \textbf{Step 4:} Calculate the optimal powers using \eqref{K_user_closed_form_interms_of_rates}. 
\end{algorithmic}
\end{small}
\end{algorithm}

\vspace{-5pt}
\subsection{Time Division Multiple Access}
\label{Binary_TDMA}
In this section we will tackle the total energy minimization problem of a system with $K'$ (completely) offloading users when TDMA is employed as the multiple access scheme. In the TDMA scheme there is only one user offloading at a time. \rev{Hence, there is no interference and the rate that each user employs when it transmits is bounded by the single-user capacity, e.g., \cite{cover2012elements}. However, since the devices are transmitting one at a time, the allowable latency of each user must include the time that the user spends waiting for the devices scheduled to transmit earlier to complete their transmission. 
Therefore, the natural transmission schedule is in the order of increasing values of the transmission latency $\tilde{L}_k$, which was defined after \eqref{binary_main_K_user_MAC}. Without loss of generality we can order the users so that $\tilde{L}_1\le \tilde{L}_2 \le \dots \le \tilde{L}_{K'}$, and in that case the waiting time of user $k$ can be written as $t_{\text{w}_k} = \sum_{i=1}^{k-1} t_{\text{UL}_i}= \sum_{i=1}^{k-1}T_s  \bigl(\tfrac{B_{i}}{R_{i}}\bigr)$.}
The device energy minimization problem in the TDMA case can then be written as
\vspace{-5pt}
\begin{subequations}
\label{binary_TDMA_prob_main_K_user}
\begin{align}
\label{binary_TDMA_prob_main_K_user_a}
\min_{\substack{\{R_{k}\}, \{P_{k}\}}}  &\textstyle \sum_{k \in \mathcal{S}'}  \tfrac{B_k}{R_{k}}P_{k} \\ 
\label{binary_TDMA_prob_main_K_user_b}
 \text{s.t.} \quad & \textstyle{\sum_{i=1}^k}T_s  \bigl(\tfrac{B_{i}}{R_{i}}\bigr) \le \tilde{L}_{k}, \quad \forall k \in \mathcal{S}', \\
 \label{binary_TDMA_prob_main_K_user_c}
& 0 \le P_{k}, \quad  \forall k \in \mathcal{S}', \\ 
\label{binary_TDMA_prob_main_K_user_d}
& 0 \le R_{k} \le \log_2(1+\alpha_k P_{k}), \quad \forall k \in \mathcal{S}'.
\end{align}
\end{subequations}
For a fixed set of transmission rates $\{R_k\} $, the objective in \eqref{binary_TDMA_prob_main_K_user_a} is an increasing function of each transmission power $P_k $, and the constraints on the powers are separable \rev{(because TDMA avoids interference between the users)}. Hence, the optimal solution for the transmission power for user $k$ is, \rev{simply, the minimum power required to achieve its target transmission rate, namely, }
\vspace{-10pt}
\begin{equation}
\label{TDMA_full_power}
P_k = \tfrac{2^{R_k}-1}{\alpha_k}.
\end{equation} 
The remaining problem can be written in terms of the transmission rates as follows 
\begin{subequations}
\label{binary_TDMA_prob_main_rates}
\begin{align}
\min_{\substack{\{R_{k}\}}} \quad  &  \textstyle \sum_{k \in \mathcal{S}'}  \tfrac{B_k}{\alpha_{k}} \bigl(\tfrac{2^{R_k}-1}{R_k}\bigr) \\ 
\label{binary_TDMA_prob_main_rates_b}
 \text{s.t.} \quad &  \textstyle{\sum_{i=1}^k}T_s  \bigl(\tfrac{B_{i}}{R_{i}}\bigr) \le \tilde{L}_{k}, \quad \forall k \in \mathcal{S}', \\
 \label{binary_TDMA_prob_main_rates_c}
& 0 \le R_{k}, \quad  \forall k \in \mathcal{S}'.
\end{align}
\end{subequations}

It can be shown that the objective function in \eqref{binary_TDMA_prob_main_rates} is jointly convex in the transmission rates and hence, the optimal solution to \eqref{binary_TDMA_prob_main_rates} can be efficiently obtained. The optimal solution to \eqref{binary_TDMA_prob_main_K_user} is then the concatenation of these rates and the corresponding powers in \eqref{TDMA_full_power}.

\subsection{Binary Computational Offloading}
Now that we have obtained a closed-form optimal resource allocation for a given set of offloading users in the case of the full multiple access scheme, and a quasi-closed-form solution based on a convex optimization problem with $K'$ variables in the case of TDMA, we can tackle the ``outer'' problem of finding an optimal set of offloading users. This is a combinatorial problem, with a search space of $2^K$ possibilities, but it admits a tree structure. Therefore, in addition to the branch-and-bound algorithm for finding an optimal set of offloading users, the problem is amenable to a wide variety of lower-complexity tree-search algorithms that typically provide offloading sets with low energy consumption. As an example, we will develop a customized greedy search technique in which the search tree is (deterministically) pruned at each iteration.

\subsubsection{Greedy search algorithm}
\label{greedy_search}
To describe the proposed algorithm, we let $\mathcal{S}'$ denote the set of users that have already been chosen for offloading, and let $\mathcal{U}$ denote the set of users for which a decision as to whether or not to offload has yet to be made. We initialize the algorithm with all the users in $\mathcal{U}$ and none in $\mathcal{S}'$. The key steps in each iteration of the algorithm are an exploratory step, a deterministic pruning step, and a greedy user selection step that selects the ``best'' user to add to the offloading set (if any remain after the pruning step). These steps are summarized in steps 3, 4, and 6 in Algorithm~\ref{binary_algorithm}. In the exploration step, for each user in $\mathcal{U}$ we obtain the energy consumption of the system if that user were to be added to the set of offloading users. In the case of FullMA scheme that can be computed using the closed-form expression in Algorithm~\ref{full_algorithm} and in the case of TDMA it can be found by solving the convex optimization problem in \eqref{binary_TDMA_prob_main_rates} and using the expression in \eqref{TDMA_full_power}. In the pruning step we remove from $\mathcal{U}$ all those users for whom the exploration step revealed that (at this iteration) offloading would incur more energy consumption than local computation. These users can be ``safely'' removed, because in subsequent iterations there will be more users offloading and hence the energy required by any individual user to offload their task does not decrease as the iterations progress. In the greedy user selection for offloading step we select the user for which offloading offers the greatest reduction in the energy consumption of the system.

To analyze the computational effort required by the algorithm, let $Q^{(i)}$ denote the cardinality of the set $\mathcal{U}$ at the beginning of the $i^{\text{th}}$ iteration; i.e., at Step 3. At each iteration of the algorithm, the exploration step involves the solution of $Q^{(i)}$ complete offloading problems (Algorithm~\ref{full_algorithm} for full multiple access scheme or \eqref{binary_TDMA_prob_main_rates} then \eqref{TDMA_full_power} for TDMA scheme). The combination of the pruning and greedy selection steps requires $Q^{(i)}$ comparisons. At iteration $i$, there are $i$ users in $\mathcal{S}'$ and hence, in the full multiple access case the cost of each complete offloading problem in Step 3 is $O(i \log i)$. Hence, the computational cost of Algorithm~\ref{binary_algorithm} in the full multiple access case is dominated by a term that is $O\bigl(\textstyle{\sum_i} Q^{(i)} i \log i \bigr)$. In the worst case, no users are pruned in Step 4, so $Q^{(i)} \le K-i+1$ and hence the computation cost is at most $O\bigl(\textstyle{\sum_{i=1}^K}(K-i+1) i \log i \bigr)$. A loose upper bound for the argument of that expression is $K^3 \log K$. In our numerical results in Section~\ref{num_results} we will show that the proposed search strategy produces solutions that typically provide near optimal energy consumption and that it does so at a low computational cost. 

We remark that an alternative greedy-based algorithm to solve the energy minimization problem of a binary offloading system was developed in \cite{ding_NOMA_arxiv} for systems in which all the users have the same latency constraint. The greedy choice at each iteration in that algorithm is similar to that in Algorithm~\ref{binary_algorithm};  i.e., at each iteration a user which results in the maximum reduction of the total energy consumption is added to the set of offloading users. However, as we will illustrate in Section~\ref{num_results} the computational cost of the greedy algorithm in \cite{ding_NOMA_arxiv} is significantly higher than that of Algorithm~\ref{binary_algorithm}. This is mainly due to the fact that each component of the equivalent to Step 2 of Algorithm~\ref{binary_algorithm} involves solving a problem using the ellipsoid algorithm. Using analysis similar to that in the previous paragraph, that results in a computational cost that is $O(K^5)$. In contrast, Algorithm~\ref{binary_algorithm} solves the corresponding problems using the closed-form expressions in Algorithm~\ref{full_algorithm}.  The analogous analysis for Algorithm~\ref{binary_algorithm} yields a computational cost that is $O(K^3 \log K)$. 

 \begin{algorithm}[t]
  \begin{small}
\caption{: Binary Offloading Solution} 
\label{binary_algorithm}
\begin{algorithmic}
\State \textbf{Input data:} values of $\{ B_k \} $, $\{ \tilde{L}_k \}$, $\{ \alpha_k \}$, $\{\underline{E}_{\text{loc}_k}\}$, $T_s$.
\State \textbf{Step 1:} Set $\mathcal{U}= \{1, 2, \dots, K\}$, $\mathcal{S}' = \emptyset$, $E_{\text{off}}^{(0)} = 0$, $i =0$. 
\State \textbf{Step 2:} Set $\mathcal{V} = \emptyset$ and $i \leftarrow i+1$.
 \For {  each $k \in \mathcal{U}$}
\State Obtain the energy consumption of the system when user $k$ is added to the set of offloading users, $E_{\text{total}_k}^{(i)}$; i.e., perform Alg.~\ref{full_algorithm}, or solve \eqref{binary_TDMA_prob_main_rates} then \eqref{TDMA_full_power}, for $\mathcal{S}' \cup \{ k\}$.
\If {$E_{\text{off}}^{(i-1)} +\underline{E}_{\text{loc}_k}\le E_{\text{total}_k}^{(i)}$}
\State  Add user $k$ to the set of users to be pruned; i.e., $\mathcal{V} \leftarrow \mathcal{V} \cup \{ k \}$
\EndIf
\EndFor
\State \textbf{Step 3:} Prune the selected users from the tree; i.e., $\mathcal{U} \leftarrow \mathcal{U} \setminus \mathcal{V}$.
\State \textbf{Step 4:} If $\mathcal{U}= \emptyset$, terminate the algorithm.
\State \textbf{Step 5:} Select the ``best'' user by choosing 
$k^{\star} = \arg \displaystyle{\max_{k \in \mathcal{U}} }\bigl(E_{\text{off}}^{(i-1)} + \underline{E}_{\text{loc}_k}-E_{\text{total}_k}^{(i)}\bigr) $.
\State \textbf{Step 6:} Update the offloading set and the undecided set; i.e., $\mathcal{S}' \leftarrow \mathcal{S}' \cup \{ k^{\star} \}$ and $\mathcal{U} \leftarrow \mathcal{U} \setminus \{ k^{\star} \}$.
\State \textbf{Step 7:}  Update the offloading energy of the system; i.e., $E_{\text{off}}^{(i)} = E_{\text{total}_{k^{\star}}}^{(i)} $.
\State \textbf{Step 8:} If $\mathcal{U}= \emptyset$, stop. If not go to Step 3.
\end{algorithmic}
\end{small}
\end{algorithm}

Algorithm~\ref{binary_algorithm} can be modified for the case of conventional local communication architecture by replacing each $\underline{E}_{\text{loc}_k}$ by the energy required to complete the task locally using the conventional architecture, and by adjusting the initialization of the offloading set $\mathcal{S}'$ and the undecided set $\mathcal{U}$. The set $\mathcal{S}'$ is initialized with those users for which the task cannot be completed locally by the deadline, and $\mathcal{U}$ is initialized as $\{1, 2, \dots, K \} \setminus \mathcal{S}'$. The initial offloading energy $E_{\text{off}}^{(0)}$ is set to be the optimal total energy consumption of the users in the initial set $\mathcal{S}'$.

\subsubsection{Rounding-based algorithm}
\label{rounding_based}
The authors in \cite{ding_NOMA_arxiv} also proposed a binary offloading algorithm for systems with equal latencies that is based on choosing the set of offloading users by rounding the solution to the corresponding partial offloading problem, and then solving the complete offloading problem for that set. 
That rounding approach extends naturally to the formulation that we have considered. 
Furthermore, since the rounding process selects 
offloading users
by rounding the optimal value of the offloading fraction in the partial offloading problem, $\gamma^{\star}_k \in [0, 1]$, to a value in $\{0, 1\}$, there is a natural extension to randomized rounding; cf. \cite{raghavan1987randomized}. In that case, multiple candidate sets of offloading users are selected according to independent Bernoulli distributions with the probability of offloading for user $k$ being $\gamma^{\star}_k$. The hybrid scheme of deterministic and randomized rounding also arises naturally.
Our numerical results in Section~\ref{num_results} will show that although the incorporation of randomized rounding offers better performance than deterministic rounding of the partial offloading solution, the proposed greedy search over the tree of complete offloading problems offers significant reductions in 
the energy consumption,
at a computational cost that is similar to that of the deterministic rounding approach. 

\section{Partial offloading}
\label{partial_off}
Up until this point, we have considered computational tasks with tightly coupled components, which must be either totally offloaded or executed locally. If the computational tasks are divisible, the device energy consumption can be reduced by taking advantage of the parallelism between the access point and the devices. In that case, each user offloads a portion of its task
and executes the remaining portion locally. As mentioned earlier, we will focus on ``data-partitionable'' tasks \cite{wang2016mobile}. Such tasks involve a relatively simple-to-describe action being applied, independently, to multiple blocks of data. As such, the number of operations required to complete a fraction of the task is modeled as being a function of the description length \cite{wang2016mobile, munoz2015optimization, zhang2013energy}, cf. \eqref{exe_time_offload} and \eqref{loc_opt_energy}. 

\subsection {Full Multiple Access} 
\label{partial_FullMA}
In this section we will
consider
a $K$-user partial offloading system that employs a FullMA scheme. The users are assumed to adopt dynamic voltage scaling so that they can minimize their local computation energy consumption. (In this setting the local latency constraint is satisfied implicitly and $E_{\text{loc}_k}$ takes the form in \eqref{loc_opt_energy}.) \rev{Considering that in a FullMA scheme $t_{\text{w}_k} = 0, \forall k$}, the energy minimization problem is
\begin{subequations}
\label{partial_MAC_prob_main_K_user}
\begin{align}
\min_{\substack{\{R_{k}\}, \{P_{k}\}, \{\gamma_{k}\}}}  & \textstyle  \sum_{k}  \tfrac{\gamma_{k}B_k}{R_{k}}P_{k} + \tfrac{M_k}{{L}_k^2} \bigl((1-\gamma_k )B_k\bigr)^3\\ 
\label{partial_MAC_prob_main_K_user_b}
 \text{s.t.} \quad &  T_s  \bigl(\tfrac{\gamma_{k}B_{k}}{R_{k}}\bigr)+ \delta_c \gamma_k B_k \le \bar{L}_{k}, \quad \forall k, \\
  \label{partial_MAC_prob_main_K_user_c}
 & 0 \le \gamma_{k} \le 1, \quad  \forall k, \\
 \label{partial_MAC_prob_main_K_user_d}
& 0 \le P_{k}, \quad  \forall k, \\ 
\label{partial_MAC_prob_main_K_user_e}
& \rev{\{R_{k}\}_{k=1}^{K}  \in \mathcal{R}\bigl( \{P_{k}\} _{k=1}^{K}\bigr),} 
\end{align}
\end{subequations}
where $\bar{L}_k = L_k- t_{{\text{DL}}_k}$. The achievable rate region for FullMA
was described in Section~\ref{binary_fullMA}.  

Using the insights generated from a two-user case in \cite{salmani_MACO_arxiv}, it can be shown that optimal solution of the problem in \eqref{partial_MAC_prob_main_K_user} is obtained when each user utilizes its maximum allowable latency, i.e., the constraints in \eqref{partial_MAC_prob_main_K_user_b} hold with equality. Accordingly, the closed-form solution for the optimal fraction of bits offloaded by the $k^{\text{th}}$ user is
\begin{equation}
\label{MAC_gamma_opt}
\gamma_k = \tfrac{\bar{L}_k R_k}{B_k(T_s+\delta_c R_k)},
\end{equation}
and the problem in \eqref{partial_MAC_prob_main_K_user} can be reduced to 
\begin{subequations}
\label{partial_MAC_prob_main_K_user_no_gamma}
\begin{align}
\min_{\substack{\{R_{k}\}, \{P_{k}\}}}  & \textstyle  \sum_{k} \tfrac{ \bar{L}_k}{T_s+\delta_c R_k} P_{k} + \tfrac{M_k}{{L}_k^2} (B_k - \tfrac{\bar{L}_k R_k}{T_s+\delta_c R_k})^3\\ 
\label{partial_MAC_prob_main_K_user_no_gamma_b}
 \text{s.t.} \quad &  0 \le  \tfrac{\bar{L}_k R_k}{B_k(T_s+\delta_c R_k)} \le 1, \quad  \forall k, \\
 \label{partial_MAC_prob_main_K_user_no_gamma_d}
& \eqref{partial_MAC_prob_main_K_user_d}, \eqref{partial_MAC_prob_main_K_user_e},
\end{align}
\end{subequations}
where \eqref{partial_MAC_prob_main_K_user_no_gamma_b} results from the constraints 
in \eqref{partial_MAC_prob_main_K_user_c}. 
The problem in \eqref{partial_MAC_prob_main_K_user_no_gamma} can be decomposed as
\begin{eqnarray}
\label{partial_main_K_user_MAC_decompose}
& \hspace{-40pt}  \displaystyle{\min_{ \{R_{k}\}}}   & \hspace{5pt} \displaystyle{\min_{\{P_{k}\}}} \hspace{5pt}  \textstyle \sum_{k} \tfrac{ \bar{L}_k}{T_s+\delta_c R_k} P_{k} \\ 
&\quad \quad \text{s.t.}  \hspace{5pt}   \eqref{partial_MAC_prob_main_K_user_e}, \eqref{partial_MAC_prob_main_K_user_no_gamma_b},  &\hspace{10pt} \text{s.t.} \hspace{10pt}   \eqref{partial_MAC_prob_main_K_user_d}, \eqref{partial_MAC_prob_main_K_user_e}. \nonumber
\end{eqnarray}
For a given set of transmission rates, the objective function in \eqref{partial_main_K_user_MAC_decompose} has a structure that is analogous to that of the objective in \eqref{binary_main_K_user_MAC_decompose}. Hence, if the permutation $\pi$ is defined such that 
\begin{equation}
\label{rho_order_partial}
\rho'_{\pi(K)} \le \rho'_{\pi({K-1})} \le \dots \le \rho'_{\pi(1)},
\end{equation} 
where 
\begin{equation}
\label{rho_prime_eq}
\rho'_k = \tfrac{\bar{L}_k}{\alpha_k (T_s+\delta_c R_k)},
\end{equation}
the following closed-form optimal solution for the transmission powers can be obtained
\begin{equation}
\label{partial_K_user_closed_form_interms_of_rates}
 P_{\pi(k)} =  (\tfrac{2^{R_{\pi(k)}}-1}{ \alpha_{\pi(k)}}) 2^{\sum_{j=1}^{k-1} R_{{\pi(j)}} }.
\end{equation} 
As in the complete offloading case, the ordering in \eqref{rho_order_partial} also specifies the decoding order that enables the rates that will be found in \eqref{partial_MAC_prob_main_K_user_no_power} to be achieved using successive decoding. 

Now the outer optimization problem in \eqref{partial_main_K_user_MAC_decompose} becomes
\begin{subequations}
\label{partial_MAC_prob_main_K_user_no_power}
\begin{align} 
\min_{\substack{\{R_{k}\}}} \quad  & \textstyle  \sum_{k} \tfrac{ \bar{L}_{\pi(k)}}{\alpha_{\pi(k)}} (\tfrac{2^{R_{\pi(k)}}-1}{T_s+\delta_c R_{\pi(k)} }) 2^{\sum_{j=1}^{k-1} R_{{\pi(j)}} } +  \textstyle  \sum_{k} \tfrac{M_k}{{L}_k^2} (B_k -\tfrac{\bar{L}_k R_k}{T_s+\delta_c R_k})^3\\ 
\label{partial_MAC_prob_main_K_user_no_power_b}
 \text{s.t.} \quad &  0 \le  \tfrac{\bar{L}_k R_k}{B_k(T_s+\delta_c R_k)} \le 1, \quad  \forall k.
\end{align}
\end{subequations}
We have shown in Appendix~\ref{MAC_partial_quasiconvexity} 
that the objective function in \eqref{partial_MAC_prob_main_K_user_no_power} is quasi-convex  in terms of each $R_k$ when the other transmission rates are fixed. In addition, the constraints on the transmission rates are separable. Therefore, the coordinate descent algorithm can be employed to find a stationary solution for the transmission rates in \eqref{partial_MAC_prob_main_K_user_no_power}; e.g., \cite[Theorem 1]{hong2016unified}. 

Using the obtained solutions for the transmission rates, we can update the values of the $\rho'_k$s in \eqref{rho_prime_eq} and consequently the optimal values of the transmission powers in \eqref{partial_K_user_closed_form_interms_of_rates}. By substituting the updated transmission powers into the problem in \eqref{partial_MAC_prob_main_K_user_no_power}, updated solutions for the transmission rates can be achieved. The resulting iterative algorithm is summarized in Algorithm~\ref{main_algorithm_partial}. \rev{The computation cost of each iteration of Algorithm~\ref{main_algorithm_partial} is dominated by the ordering in Step~3 of the algorithm, the complexity of which is $O(K \log K)$. }
While the development of a formal convergence analysis of Algorithm~\ref{main_algorithm_partial} remains a work in progress, in our numerical experience, some of which is reported in Section~\ref{num_results}, the algorithm always converged \rev{quite fast; typically in 2--5 iterations and in no more than 10 iterations}. 
 \begin{algorithm}[t]
 \begin{small}
\caption{: Iterative algorithm for \eqref{partial_MAC_prob_main_K_user}} 
\label{main_algorithm_partial}
\begin{algorithmic}
\State \textbf{Input data:} $\{B_k\} $, $\{ \bar{L}_k \}$, $\{ M_k \}$, $\{ \alpha_k \}$, $T_s$, and $\delta_c$. 
\State \textbf{Step 1:} Initialize $\{R_k\}$ so that \eqref{partial_MAC_prob_main_K_user_d} and \eqref{partial_MAC_prob_main_K_user_no_power_b} are satisfied.
\State \textbf{Step 2:} Calculate the optimal $\{ \gamma_k \}$ using \eqref{MAC_gamma_opt}.
\State \textbf{Step 3:} Calculate $\{ \rho'_k \}$ using \eqref{rho_prime_eq}.
\State \textbf{Step 4:} Order $\{ \rho'_k \}$ according to \eqref{rho_order_partial}.
\State \textbf{Step 5:} Calculate the optimal powers using \eqref{partial_K_user_closed_form_interms_of_rates}. 
\State \textbf{Step 6:} Find a stationary point of the problem in \eqref{partial_MAC_prob_main_K_user_no_power}. 
\State \textbf{Step 7:} If the convergence criterion has been satisfied terminate the algorithm. Otherwise return to Step 2.
\end{algorithmic}
\end{small}
\end{algorithm}

\subsection{Time Division Multiple Access}
\label{partial_TDMA}
For a TDMA-based partial offloading system, \rev{if we order the users such that
$\bar{L}_1 \le \bar{L}_2 \le \cdots \le \bar{L}_K$, where $\bar{L}_k$ was defined after \eqref{partial_MAC_prob_main_K_user}, then, analogous to the binary offloading case, the waiting time for user $k$ is $t_{\text{w}_k} = \sum_{i=1}^{k-1}T_s  \bigl(\tfrac{\gamma_{i}B_{i}}{R_{i}}\bigr)$,}
and the device energy minimization problem for the optimized dynamic voltage scaling architecture can be written as 
\begin{subequations}
\label{partial_TDMA_prob_main_K_user2}
\begin{align}
\label{partial_TDMA_prob_main_K_user2_a}
\min_{\substack{\{R_{k}\}, \{P_{k}\},  \{\gamma_{k}\}}}  & \textstyle  \sum_{k} \tfrac{\gamma_{k}B_k}{R_{k}}P_{k} + \tfrac{M_k}{{L}_k^2} \bigl((1- \gamma_{k})B_k\bigr)^3 \\ 
\label{prob_formula_main_K_user2_b}
 \text{s.t.} \quad &  \textstyle{\sum_{i=1}^k}T_s  \bigl(\tfrac{\gamma_{i}B_{i}}{R_{i}}\bigr)+ \delta_c \gamma_k B_k \le \bar{L}_{k}, \quad \forall k, \\
  \label{partial_TDMA_prob_main_K_user2_c}
 & 0 \le \gamma_{k} \le 1, \quad  \forall k, \\
 \label{partial_TDMA_prob_main_K_user2_d}
& 0 \le P_{k}, \quad  \forall k, \\ 
\label{partial_TDMA_prob_main_K_user2_e}
& 0 \le R_{k} \le \log_2(1+\alpha_k P_{k}), \quad \forall k.
\end{align}
\end{subequations}

For a given set of $(\{R_k\}, \{\gamma_k\})$ the objective 
is increasing in each $P_k$, and the constraints on the powers
are separable. Hence, the optimal powers are
the minimum feasible values; i.e., 
$P_k = \tfrac{2^{R_k}-1}{\alpha_k}$. If we let $B'_k = \gamma_k B_k$ and $t_k = \tfrac{B'_k}{R_k}$ denote the offloaded portion of the computational task for the $k^{th}$ user, and the time it takes to offload that portion to the access point, respectively, the problem in \eqref{partial_TDMA_prob_main_K_user2} 
can then be written as 
\begin{subequations}
\label{partial_TDMA_prob_main_K_user_no_power}
\begin{align}
\min_{\substack{\{B'_{k}\}, \{t_{k}\}}}  & \textstyle  \sum_{k} t_k \tfrac{2^{B'_k/t_k}-1}{\alpha_k} + \tfrac{M_k}{{L}_k^2} (B_k- B'_k)^3 \\ 
\label{prob_formula_main_K_user2_b}
 \text{s.t.} \quad &  \textstyle{\sum_{i=1}^k}T_s  t_i+ \delta_c B'_k \le \bar{L}_{k}, \quad \forall k, \\
  \label{partial_TDMA_prob_main_K_user_no_power_b}
 & 0 \le B'_k \le B_k, \quad  \forall k, \\
 \label{partial_TDMA_prob_main_K_user_no_power_c}
& 0 \le t_{k}, \quad  \forall k.  
\end{align}
\end{subequations}
It is shown in Appendix~\ref{partial_TDMA_jointlyconvex} 
that this problem is jointly convex in $\{B'_k\}$ and $\{t_k\}$ and hence, the optimal solution of the problem can be efficiently obtained. 

\section{Numerical Results}
\label{num_results}
In this section we will evaluate the performance of the proposed energy minimization algorithms in both binary offloading and partial offloading scenarios, using either a full multiple access scheme (FullMA) or TDMA. We will compare the performance and computational cost of the proposed algorithms to those in  \cite{ding_NOMA_arxiv} and \cite{you2017energy}.
The approach in \cite{ding_NOMA_arxiv} is a ``full multiple access'' approach, but is constrained to the case in which the latencies of the users are the same. Furthermore, the algorithm in  \cite{ding_NOMA_arxiv} does not exploit as much of the algebraic structure of the problem as our algorithm and hence its computational cost grows more quickly than that of the proposed algorithm; see the discussion in Section~\ref{greedy_search}. The approach in \cite{you2017energy} tackles the energy minimization problem for partial offloading in the TDMA case. Like the approach in \cite{ding_NOMA_arxiv}, it is also constrained to the case in which the latencies of all users are the same. The approach in \cite{you2017energy} is developed for conventional local computing architectures whereas the proposed approaches and those in \cite{ding_NOMA_arxiv} are developed for the dynamic voltage scaling architecture. 

We will consider a cell of radius 1,000m over which the users are uniformly distributed. The symbol interval is $T_s=10^{-6} \text{s}$ and we consider a 
slowly fading channel model with a path-loss exponent of 3.7 and independent Rayleigh distributed small-scale fading. The receiver noise variance is set to $\sigma^2 = 10^{-13}$.  The energy consumption in each experiment is averaged over 100 channel realizations. We assume that the time it takes to download the results to the mobile users is equal for all the users, $t_{\text{DL}_k} =0.2$s. 

\subsection{Binary Computation Offloading}
\label{sim_binary}
In the first phase of our numerical experiments we will \rev{consider the case where the users seek to complete indivisible computational tasks}, and hence they should either offload their task or complete it locally. 
We will begin by considering a four-user system in which the users latencies are different, $[L_1, L_2, L_3, L_4]= [1.2, 1.5, 1.8, 2.5]$s, and we will examine the energy consumption of FullMA and TDMA-based binary offloading systems as the (different) description lengths of the tasks grow (in proportion); $[B_1, B_2, B_3, B_4]= \zeta \times [2, 1, 3, 4] \times 10^{6}$ bits. \rev{In order to model the optimized energy consumption of local execution in each user, while meeting its latency constraint, we set $\underline{E}_{\text{loc}_k} = \tfrac{M_k}{{L}_k^2} B_k^3$ (see \eqref{binary_DVS}), and to be consistent with the measurements in \cite{miettinen2010energy}, we set $M_k= 10^{-19}$ \cite{zhang2013energy, wang2016mobile}. } We apply the proposed greedy algorithm (Algorithm~\ref{binary_algorithm}) to both a FullMA scheme and the TDMA scheme to find a good set of offloading users and the corresponding power and rate allocation. We will compare the energy consumption of these schemes to that of schemes in which the offloading set is chosen by deterministic rounding of the solution of the corresponding partial offloading problem, and to a scheme that selects the best solution from the deterministically rounded case and $(K-1)$ randomized roundings; see Section~\ref{rounding_based}. In the case of FullMA we compare the performance and the computational cost of the proposed algorithm with those of the binary offloading algorithm proposed in \cite{ding_NOMA_arxiv}.

\begin{figure}[t]
\centering
\includegraphics[width=1\linewidth,width=0.47\textwidth]{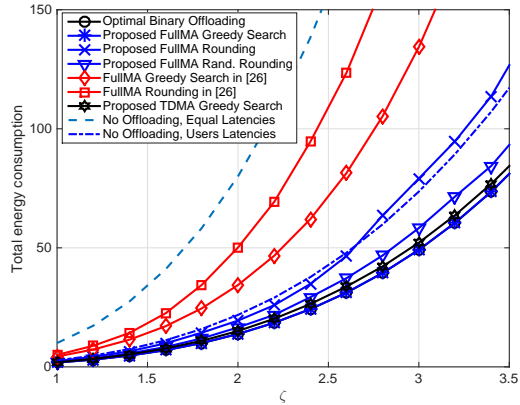} 
\caption{Average energy consumption of a binary offloading system with four users with different latency constraints versus the parameter that defines the required number of bits to describe the users' tasks.} 
\label{Binary_EvsB}
\end{figure}

\begin{figure}
\centering
\includegraphics[width=1\linewidth,width=0.47\textwidth]{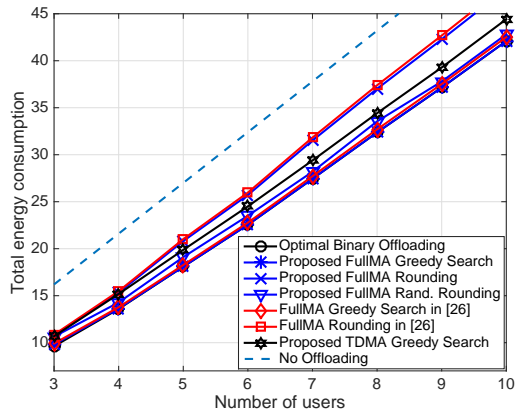} 
\caption{Average energy consumption of a binary offloading system, in which the users' tasks have the same latency constraints, for different number of users.} 
\label{Binary_EvsU}
\end{figure}

Fig.~\ref{Binary_EvsB} plots the average energy consumption of the four-user system as the problem sizes grow. Our first observation is that the proposed greedy search algorithm to find a set of offloading users provides close-to-optimal performance for both FullMA and TDMA, and significantly better performance than the deterministic rounding approach. In this setting, the optimized TDMA scheme performs quite well, but in other scenarios that we will consider (Figs~\ref{Binary_EvsU}, \ref{partial_EvsB}, and \ref{partial_EvsU}) an optimized FullMA scheme enables a significantly larger reduction in the energy consumption.

It can be seen from Fig.~\ref{Binary_EvsB} that utilizing the maximum available latencies of the users enables the proposed algorithm to substantially reduce the energy consumption compared to the algorithm in \cite{ding_NOMA_arxiv}, in which the users are assumed to have the same latency constraints. The performance gap increases quite quickly as the sizes of the problems increase. In the ``No Offloading'' approach in Fig.~\ref{Binary_EvsB}, the users complete their tasks locally employing the dynamic voltage scaling approach, by which they can minimize the local energy consumption 
subject to their latency constraints.
Interestingly, the energy consumption when all users complete their tasks locally using the maximum available latency is substantially less than that of the latency-equal algorithm proposed in \cite{ding_NOMA_arxiv} and the case in which the offloading set is chosen by deterministically rounding the solution to the partial offloading problem with different latencies.

In order to compare the computational costs of the proposed FullMA  algorithm with that in \cite{ding_NOMA_arxiv}, Table~\ref{table_binary_cost} provides the average CPU times. These times are essentially independent of the description length of the tasks. All the algorithms were coded in \textsc{Matlab}, with similar diligence paid to the efficiency of the programs. The convex optimization subproblems in the method in \cite{ding_NOMA_arxiv} were solved using \textsc{SDPT3} \cite{toh1999sdpt3} through the CVX interface \cite{grant2008cvx}. The CPU times were evaluated on a MacBook Pro with a Core i5 processor running at 3.1GHz, and 8GB of RAM. It can be seen that the closed-form optimal solution that we have obtained for any given set of offloading users significantly reduces the computational cost of our proposed algorithm in comparison to the algorithm in \cite{ding_NOMA_arxiv}. As discussed in Section~\ref{greedy_search}, the main reason for such a significant computational cost reduction is that at each iteration of the proposed algorithm the optimal closed-form solution for a given set of offloading users is obtained with the cost of order $O(K \log K)$, while at each iteration of the algorithm proposed in \cite{ding_NOMA_arxiv} an optimization problem needs to be solved by employing the ellipsoid method which involves matrix inversion with the cost of order $O(K^3)$. (As suggested in \cite{ding_NOMA_arxiv}, for the ellipsoid method we employed the approach in \cite{boyd2008ellipsoid}, and we chose a termination criterion of $\epsilon = 10^{-3}$.)

\begin {table}
\caption {Average CPU times required for the proposed algorithms and the algorithms in \cite{ding_NOMA_arxiv} for a four-user binary offloading system that employs a full multiple access scheme.} 
\label{table_binary_cost}
\vspace{-10pt}
\begin{center}
\begin{tabular} {llr}  
\toprule
\textbf{Algorithm}    & \textbf{Average CPU time (sec)}   \\
\midrule
Proposed FullMA Greedy Search       &  $4.6\times 10^{-5}$         \\ 
\midrule
Proposed FullMA Rounding     & $4.0 \times 10^{-5}$ \\
\midrule
FullMA Greedy Search  in \cite{ding_NOMA_arxiv}   & $1.7 \times 10^{3}$        \\
\midrule
FullMA Rounding in \cite{ding_NOMA_arxiv}    & $0.2 \times 10^{3}$    \\
\bottomrule
\end{tabular}
\end{center}
\vspace{-30pt}
\end {table}

In our next numerical experiment for the binary offloading case, we examine the total energy consumption as the number of users increases. In this experiment we consider a scenario in which all the users have equal problem sizes and the same latency constraints. In particular, we set $B_k = 6\times10^{6}$ bits and $L_k = 2$s. As in the previous experiment, the ``randomized rounding'' scheme refers to the selection of the best solution from offloading sets that are generated by a deterministic rounding of the partial offloading solution and $(K-1)$ randomized roundings.

\begin{figure}
\centering
\includegraphics[width=1\linewidth,width=0.47\textwidth]{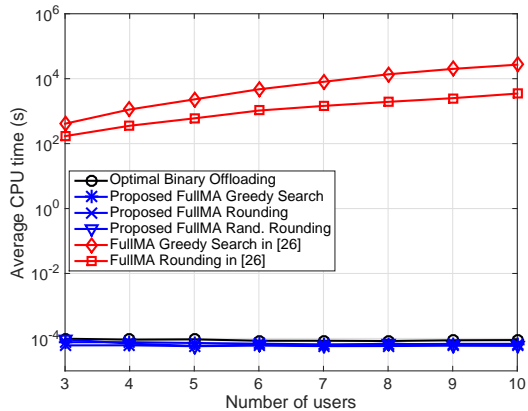} 
\caption{Average CPU time required for the proposed algorithm and the algorithm in \cite{ding_NOMA_arxiv} for different number of users when a full multiple access scheme is employed in binary offloading case.} 
\label{Binary_ComplexvsU}
\end{figure}

In Fig.~\ref{Binary_EvsU} we present the average energy consumption versus the number of users. In this setting all of the considered methods provide a significant reduction in the energy consumption over the No Offloading case. In the case that a FullMA scheme is employed, it can be seen that since the latency constraints of all the users are equal, the algorithm in \cite{ding_NOMA_arxiv} can achieve the same performance as our proposed algorithm, for both greedy search and rounding approaches. However, Fig.~\ref{Binary_ComplexvsU} indicates that 
the computational cost of the proposed algorithm is significantly less than that of the algorithm in \cite{ding_NOMA_arxiv}. Fig.~\ref{Binary_EvsU} also shows that by using the full capabilities of the channel, a FullMA scheme together with the proposed greedy search method can reduce the total energy consumption compared to the TDMA scheme with the same greedy approach. 

\subsection{Partial Computation Offloading}
\label{sim_partial}
\begin{figure}
\centering
\includegraphics[width=1\linewidth,width=0.47\textwidth]{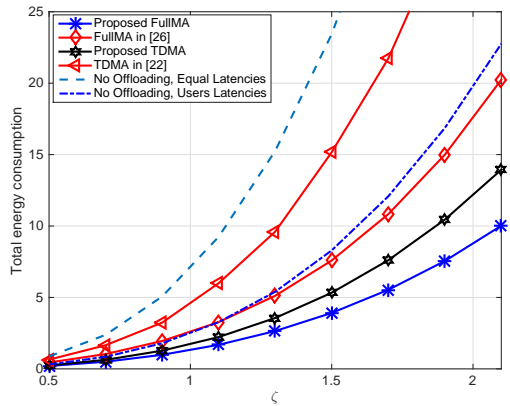} 
\caption{Average energy consumption of a four-user partial offloading system with different latency constraints versus the coefficient that defines the description length of the tasks.} 
\label{partial_EvsB}
\end{figure}

\rev{In the second phase of our numerical analysis we consider partial offloading of ``data-partitionable'' divisible computational tasks
for which the (optimal) local energy consumption can be modeled as a function of number of bits, see \eqref{loc_opt_energy}, with $M_k= 10^{-19}$; \cite{zhang2013energy, wang2016mobile}. 
To make fair comparisons with the conventional local computational architecture considered in \cite{you2017energy}, we consider problems that require 1,000 computational cycles per bit, and we set the local computing energy per cycle for each user in such a way that that user is able to complete its computational task locally within its latency constraint.}
We first examine the energy consumption of a four-user system \rev{analogous to that in Section~\ref{sim_binary}, in which the latencies are $[L_1, L_2, L_3, L_4]= [1.2, 1.5, 1.8, 2.5]$s and the description lengths grow as $[B_1, B_2, B_3, B_4]= \zeta \times [2, 1, 3, 4] \times 10^{6}$ bits.}
Fig.~\ref{partial_EvsB} plots the total energy consumption as the problem sizes grow.
It can be seen that our proposed algorithms, which benefit from the maximum available latency of each user, achieve substantially lower energy consumption than the existing techniques. Indeed, it can be seen that in the TDMA case, the energy consumption of the proposed algorithm is lower than that of the algorithm in \cite{you2017energy}, and the performance gap increases as the number of bits increases. That is because in the proposed algorithm the users not only utilize their maximum available deadline to complete their tasks, they also employ dynamic voltage scaling which minimizes the local energy consumption. The energy consumptions in Fig.~\ref{partial_EvsB} and the computational costs in Table~\ref{table_partial_cost} indicate that in the FullMA case the proposed algorithm can achieve significantly lower energy consumption than the algorithm in \cite{ ding_NOMA_arxiv}, and does so at much lower computational cost. 
Fig.~\ref{partial_EvsB} also exhibits the impact of the multiple access scheme. Using a FullMA scheme substantially reduces the total energy consumption over TDMA.

\begin{figure}
\centering
\includegraphics[width=1\linewidth,width=0.47\textwidth]{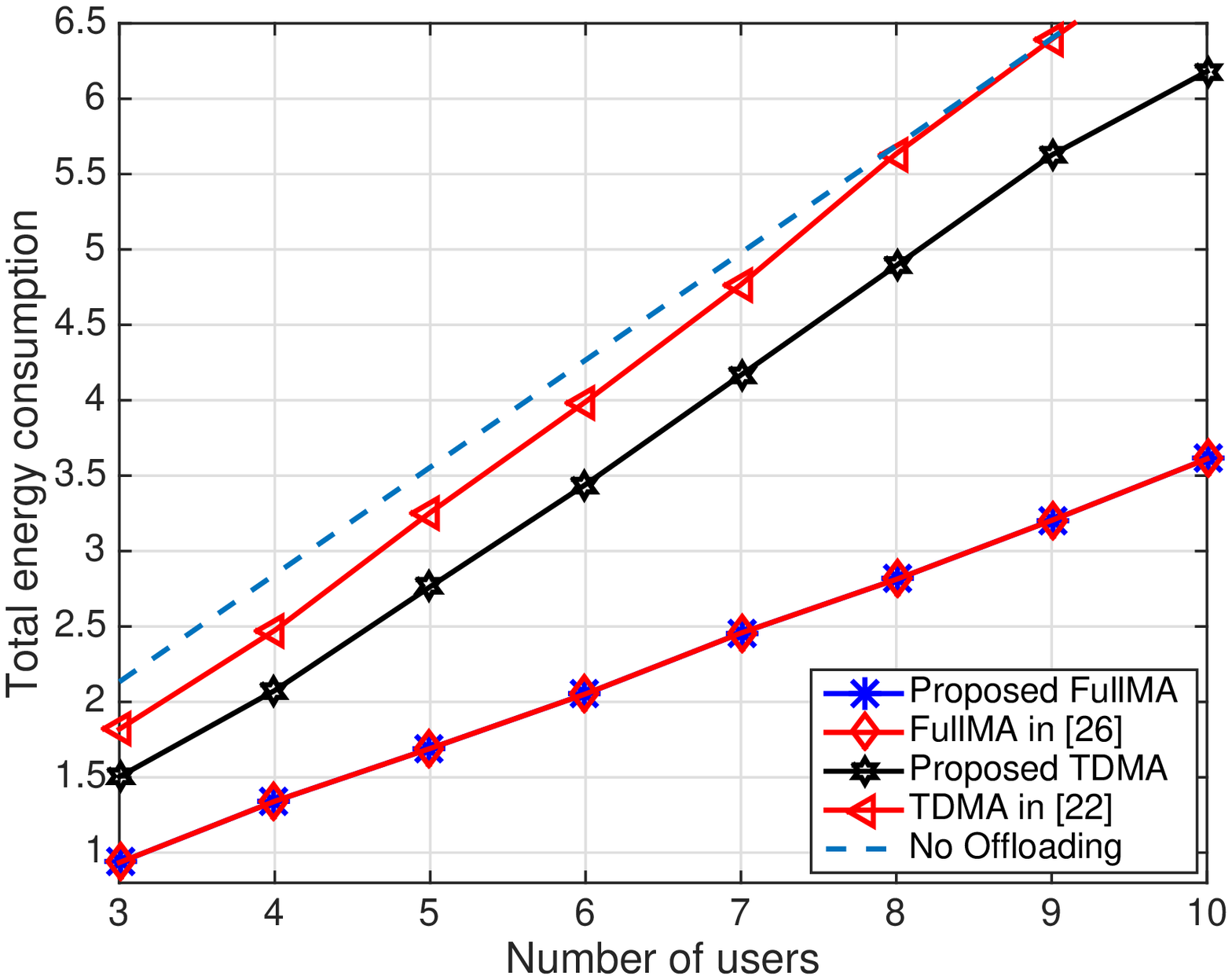} 
\caption{Average energy consumption of a partial offloading system, in which the users' tasks have the same latency constraints, for different number of users.} 
\label{partial_EvsU}
\end{figure}

\begin{figure}
\centering
\includegraphics[width=1\linewidth,width=0.47\textwidth]{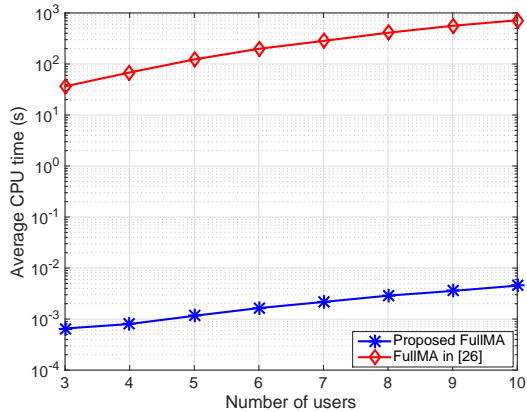} 
\caption{Average CPU time required for the proposed algorithm and the algorithm in \cite{ding_NOMA_arxiv} for different number of users when the full multiple access scheme is employed in partial offloading case.} 
\label{partial_complexvsU}
\end{figure}

In our final numerical experiment we examine the energy consumption as the number of users increases for a partial offloading system with equal problem sizes and the same latency constraints. We set $B_k = 4 \times 10^{6}$ bits and $L_k = 2$s. Fig.~\ref{partial_EvsU}, like Fig.~\ref{partial_EvsB}, shows that using the full capabilities of the channel enables the users to complete their computational tasks with significantly less energy consumption compared to TDMA. In the FullMA case, it can be seen in Fig.~\ref{partial_EvsU} that because the latencies of the users are equal, the algorithm in \cite{ ding_NOMA_arxiv} can achieve the same performance as our proposed algorithm. However, as it can be seen in Fig.~\ref{partial_complexvsU} the computational cost of the proposed algorithm is much lower. We can also see in Fig.~\ref{partial_EvsU} that when TDMA is employed, the proposed algorithm achieves noticeably lower energy consumption than that in \cite{you2017energy} despite the fact that the latencies are equal in this scenario. The reason for this is that the proposed algorithm is for systems with dynamic voltage scaling, which enables the users to minimize the energy that they expend on the portion of the task that is computed locally.

\begin {table}
\caption {Average CPU times for the proposed algorithm and the algorithm in \cite{ding_NOMA_arxiv} for a four-user FullMA partial offloading system.} 
\label{table_partial_cost}
\vspace{-15pt}
\begin{center}
\begin{tabular} {llr}  
\toprule
\textbf{Algorithm}    & \textbf{Average CPU time (sec)}   \\
\midrule
Proposed FullMA      &  $4.1 \times 10^{-3}$         \\
\midrule
FullMA in \cite{ding_NOMA_arxiv}   & $1.9 \times 10^2$        \\
\bottomrule
\end{tabular}
\end{center}
\vspace{-30pt}
\end {table}

\section{Conclusion}
In this work, we have considered the problem of optimal uplink resource allocation in a $K$-user offloading system. In the binary offloading case, we obtained the optimal energy consumption of a given set of offloading users under a full multiple access scheme and under the TDMA scheme, and then we proposed a customized greedy search algorithm to find a set of offloading users with close-to-optimal energy consumption. In the partial offloading case, the energy minimization problem was tackled by proposing a low-complexity algorithm for a stationary solution in the full multiple access case and by finding the optimal solution of a convex optimization problem when TDMA is employed. Our strategy to decompose the optimization problem and to find the optimal values of some variables in terms of the others enabled us to significantly reduce the computational cost of our proposed algorithms compared to the existing algorithms in this area. 

While the proposed resource allocation algorithms have significant advantages over the existing algorithms, like the existing algorithms they have been based on a single time slot for communication. Recent work on the two-user case \cite{salmani_MACO_arxiv} suggests that a further reduction in the energy consumption can be obtained by adopting a time-slotted structure in which different groups of users transmit in each time slot. One avenue for future work is the development of efficient resource allocation algorithms for the time-slotted structure.

\appendices

\section{Exploiting the Polymatroid Structure of the Power Feasibility Region}
\label{exploit_structure}
Given the definition of class-$\ell$ constraints in Section~\ref{binary_closed_form_power}, we will show that the candidate vertices are the result of simultaneous satisfaction with equality of a set of $K$ constraints in \eqref{binary_main_K_user_MAC_d} such that there is at most one constraint from each class. 
To do so, let us assume that there are two constraints in \eqref{binary_main_K_user_MAC_d} that are satisfied with equality, namely $C_1$ and $C_2$, both of which belong to class-$c$. If $\mathcal{M}_{\text{com}}$ denotes the set of users that are present in both $C_1$ and $C_2$, and if $\mathcal{M}_{c_1}$ and $\mathcal{M}_{c_2}$ denote the set of users that are participating only in $C_1$ and $C_2$ respectively, we can write 
\begin{equation} \notag
C_1  \!: \quad 2^{\bigl(\sum_{i \in \mathcal{M}_{\text{com}}} R_i + \sum_{j \in \mathcal{M}_{c_1}} R_j\bigr)} = 
\textstyle  1+ \sum_{i \in \mathcal{M}_{\text{com}}} \alpha_i P_i  +\sum_{j \in \mathcal{M}_{c_1}} \alpha_j P_j ,
\end{equation}
\begin{equation}\notag
C_2  \!: \quad 2^{\bigl(\sum_{i \in \mathcal{M}_{\text{com}}} R_i + \sum_{k \in \mathcal{M}_{c_2}} R_k\bigr)}   =  \\
\textstyle 1+ \sum_{i \in \mathcal{M}_{\text{com}}} \alpha_i P_i  +\sum_{k \in \mathcal{M}_{c_2}} \alpha_k P_k. 
\end{equation}
By adding the above two equations we have that 
\begin{multline}
\label{adding}
\textstyle
\bigl(2^{\sum_{i \in\mathcal{M}_{\text{com}}} R_i}\bigr) \bigl(2^{ \sum_{j \in \mathcal{M}_{c_1}} R_j} + 2^{\sum_{k \in \mathcal{M}_{c_2}} R_k}\bigr)
 -1 - \textstyle \sum_{i \in \mathcal{M}_{\text{com}}} \alpha_i P_i= \\
 1+ \textstyle \sum_{i \in \mathcal{M}_{\text{com}}} \alpha_i P_i+ \textstyle \sum_{j \in \mathcal{M}_{c_1}} \alpha_j P_j +\textstyle \sum_{k \in \mathcal{M}_{c_2}} \alpha_k P_k.
\end{multline}
In addition, there is a rate region constraint that includes all the users in $\mathcal{M}_{\text{com}}  \cup \mathcal{M}_{c_1} \cup \mathcal{M}_{c_2}$,
\begin{multline}
\label{total_constraint}
2^{\bigl(\sum_{i \in \mathcal{M}_{\text{com}}} R_i + \sum_{j \in \mathcal{M}_{c_1}} R_j + \sum_{k \in \mathcal{M}_{c_2}} R_k\bigr)}\le  1+
 \textstyle \sum_{i \in \mathcal{M}_{\text{com}}} \alpha_i P_i+\sum_{j \in \mathcal{M}_{c_1}} \alpha_j P_j +\sum_{k \in \mathcal{M}_{c_2}} \alpha_k P_k.
\end{multline}
The right hand side of \eqref{total_constraint} can be replaced by its equivalent term given on the left hand side of \eqref{adding}. That results in 
\begin{multline}
\label{total_constraint_2}
2^{\bigl(\sum_{i \in \mathcal{M}_{\text{com}}} R_i + \sum_{j \in \mathcal{M}_{c_1}} R_j + \sum_{k \in \mathcal{M}_{c_2}} R_k\bigr)}\le   \\
 \textstyle \bigl(2^{\sum_{i \in \mathcal{M}_{\text{com}}} R_i}\bigr) \bigl(2^{ \sum_{j \in \mathcal{M}_{c_1}} R_j} + 2^{\sum_{k \in \mathcal{M}_{c_2}} R_k}\bigr) -1 -  \textstyle \sum_{i \in \mathcal{M}_{\text{com}}} \alpha_i P_i  \le \\
  \textstyle \bigl(2^{\sum_{i \in \mathcal{M}_{\text{com}}} R_i}\bigr) \bigl( 2^{ \sum_{j \in \mathcal{M}_{c_1}} R_j} + 2^{\sum_{k \in \mathcal{M}_{c_2}} R_k} +1 \bigr),
\end{multline}
where the second inequality in \eqref{total_constraint_2} is obtained from the rate region constraint $2^{\sum_{i \in \mathcal{M}_{\text{com}}} R_i} \le 1+ \textstyle \sum_{i \in \mathcal{M}_{\text{com}}} \alpha_i P_i $. 
By factoring out the term $2^{\sum_{i \in \mathcal{M}_{\text{com}}} R_i}$, we obtain
\begin{multline*}
\bigl(2^{ \sum_{j \in \mathcal{M}_{c_1}} R_j + \sum_{k \in \mathcal{M}_{c_2}} R_k} \bigr)\le  \textstyle 2^{ \sum_{j \in \mathcal{M}_{c_1}} R_j} + 2^{\sum_{k \in \mathcal{M}_{c_2}} R_k} -1 \\
\Rightarrow 0 \le \bigl(2^{ \sum_{j \in \mathcal{M}_{c_1}} R_j} -1  \bigr) \bigl(1- 2^{ \sum_{k \in \mathcal{M}_{c_2}} R_k}   \bigr),
\end{multline*}
which is a contradiction, because of the fact that $0 \le R_i$ and hence $0 \le 2^{ \sum_{j \in \mathcal{M}_{c_1}} R_j} -1$ for any $j$. Therefore, at an optimal vertex of the inner problem in \eqref{binary_main_K_user_MAC_decompose} no more than one constraint from any class can hold with equality. 

\section{Quasi-convexity of the Objective Function in \eqref{partial_MAC_prob_main_K_user_no_power}}
\label{MAC_partial_quasiconvexity}
A function $f$ is quasi-convex if at least one of the following conditions holds \cite{boyd2004convex}: (a) $f$ is non-increasing; (b) $f$ is non-decreasing; 
(c) there is a (turning) point, $c$, such that for any $x \le c$ the function $f(x)$ is non-increasing and for any $x \ge c$ the function $f(x)$ is non-decreasing. We will show that for each $R_k$, when the other transmission rates are constant, the objective function in \eqref{partial_MAC_prob_main_K_user_no_power} will satisfy either condition (b) or condition (c). 
We begin by rewriting that objective as 
\begin{equation}
f_k =  \Lambda_k (\tfrac{2^{R_k}-1}{T_s+\delta_c R_k }) + \Omega_k 2^{R_k}  + \tfrac{M_k}{{L}_k^2} (B_k - \tfrac{\bar{L}_k R_k}{T_s+\delta_c R_k})^3,
\end{equation}
where $\Lambda_k = \tfrac{\bar{L}_k}{\alpha_k} 2^{\sum_{j=1}^{k-1} R_{j} }$ and $\Omega_k = \textstyle{\sum_{i=k+1}^K }\tfrac{ \bar{L}_i}{\alpha_i} (\tfrac{2^{R_i}-1}{T_s+\delta_c R_i })  2^{\sum_{j\neq k}^{i-1} R_{j} }$
are always positive. The derivative of $f_k$ with respect to $R_k$ can be then written as $\tfrac{df_k}{dR_k} =  \tfrac{F_r}{(T_s + \delta_c R_{k})^2}$,
where
\begin{subequations}
\label{partial_derivative_to_Rk_factorized}
\begin{align} \notag 
 F_r  =~ &  \Lambda_k \bigl(\ln 2 ~(T_s + \delta_c R_{k})~2^{R_k}   - \delta_c(2^{R_k}-1)\bigr) + \Omega_k \ln 2 ~(T_s + \delta_c R_{k})^2~ 2^{R_k}\\ \notag
&  - 3\bar{L}_k T_s \tfrac{M_k}{{L}_k^2} (B_k - \tfrac{\bar{L}_k R_k}{T_s+\delta_c R_k})^2.
\end{align}
\end{subequations}

As $\tfrac{1}{(T_s + \delta_c R_{k})^2}$ is always positive, 
to show that either condition (b) or condition (c) holds, it is sufficient to show that $F_r$ is non-decreasing.
In order to show that,
we will show that the derivative of $F_r$ with respect to $R_{k}$ is always non-negative. The derivative is  
\begin{subequations}
\label{partial_derivative_to_R11_factorized}
\begin{align} \notag 
 \tfrac{dF_r}{dR_k}  =~ &  \Lambda_k \bigl(\ln^2 2 ~(T_s + \delta_c R_{k})~2^{R_k} \bigr) + \Omega_k \ln 2  \bigl(  \ln 2 ~(T_s + \delta_c R_{k})^2 + 2 \delta_c (T_s + \delta_c R_{k})  \bigr) ~ 2^{R_k}\\ \notag
&  + 6 \bar{L}_k T_s \tfrac{M_k}{{L}_k^2} \bigl(B_k - \tfrac{\bar{L}_k R_k}{T_s+\delta_c R_k}) (\tfrac{\bar{L}_k T_s}{(T_s+\delta_c R_k)^2}\bigr).
\end{align}
\end{subequations}
Considering the constraint in \eqref{partial_MAC_prob_main_K_user_no_power_b}, $\tfrac{dF_r}{dR_{k}}$ is a summation of non-negative terms. Hence, $\tfrac{dF_r}{dR_{k}}$ is non-negative, and hence $F_r$ is non-decreasing. 
\section{Joint Convexity of the Objective Function in \eqref{partial_TDMA_prob_main_K_user_no_power}}
\label{partial_TDMA_jointlyconvex}
In order to show that the objective function in \eqref{partial_TDMA_prob_main_K_user_no_power} is jointly convex in terms of $\{B'_k\}$ and $\{t_k\}$ we will show that the Hessain matrix of the objective is positive semidefinite. The first and the second derivatives of the objective, $f(\cdot)$, with respect to each of the $B'_k$'s and $t_k$'s are 
\begin{subequations}
\begin{gather}
 \tfrac{\partial f}{\partial t_k} = \tfrac{1}{\alpha_k} \Bigl(2^{B'_k/t_k}-1- \ln 2~   \bigr(\tfrac{B'_k}{t_k}\bigr) ~2^{B'_k/t_k} \Bigr), \quad \tfrac{\partial f}{\partial B'_k} = \tfrac{  \ln 2}{\alpha_k} ~2^{B'_k/t_k}  - \tfrac{3M_k}{L_k^2} (B_k -B'_k)^2, \\ 
\label{f_t_2}
 \tfrac{\partial^2 f}{\partial t_k^2} = \tfrac{1}{\alpha_k} \Bigl( \ln^2 2~ \bigl( \tfrac{{B'_k}^2}{{t_k}^3} \bigr) ~2^{B'_k/t_k}   \Bigr), \quad \tfrac{\partial^2 f}{\partial {B'_k}^2} = \tfrac{  \ln^2 2}{\alpha_k t_k} ~2^{B'_k/t_k}  + \tfrac{6M_k}{L_k^2} (B_k -B'_k), \\
\label{f_tb_2}
 \tfrac{\partial^2 f}{\partial t_k \partial B'_k} = \tfrac{1}{\alpha_k} \Bigl( -\ln^2 2~ \bigl( \tfrac{{B'_k}}{{t_k}^2} \bigr) ~2^{B'_k/t_k}   \Bigr),
\end{gather}
\end{subequations}
and for $j \neq k$, $\tfrac{\partial^2 f}{\partial t_j \partial t_k}$, $\tfrac{\partial^2 f}{\partial B'_j \partial B'_k}$, and $\tfrac{\partial^2 f}{\partial t_j \partial B'_k}$ are all zero.
The Hessian matrix $H \in \mathbb{R}^{2K \times 2K}$ can be constructed as the block matrix 
$H= \left[ \begin{smallmatrix}  H_{11}   & H_{12}  \\    H_{12}^T  & H_{22}
\end{smallmatrix}\right]$,
where $H_{11}$, $H_{22}$, and $H_{12}$ are diagonal matrices with $i^{\text{th}}$ diagonal elements, $\tfrac{\partial^2 f}{\partial {t_i}^2}$, $\tfrac{\partial^2 f}{\partial {B'_i}^2}$, and $\tfrac{\partial^2 f}{\partial {t_i} \partial {B'_i}}$, respectively.
It can be seen from \eqref{f_t_2} that all the elements of $H_{11}$ are positive, and hence $H_{11} \succ 0$. Moreover, using \eqref{f_t_2} and \eqref{f_tb_2}, we can show that the inequality $\tfrac{\partial^2 f}{\partial {B'_k}^2} \times  \tfrac{\partial^2 f}{\partial t_k^2} - (\tfrac{\partial^2 f}{\partial t_k \partial B'_k})^2 \ge 0$
holds for each user. These inequalities, together with the fact that the sub-blocks of the matrix $H$ are diagonal matrices, illustrate that the Schur complement of the matrix $H_{11}$ in $H$ is positive semidefinite; i.e., $H_{22} - H_{12}^T H_{11}^{-1} H_{12} \succeq 0$. Hence, the matrix $H$ is positive semidefinite \cite{boyd2004convex} and the objective function is jointly convex.

\bibliographystyle{IEEEtran}
\bibliography{references}

\end{document}